\documentclass[apl,preprint,showpacs,amsmath,amssymb,superscriptaddress,endfloats]{revtex4}

\usepackage{graphicx}% Include figure files
\usepackage{dcolumn}% Align table columns on decimal point
\usepackage{bm}% bold math

\begin{document}

\title{Current induced resistance change of magnetic tunnel junctions with ultra-thin MgO tunnel barriers}

\author{Patryk Krzysteczko}
\email{patryk@physik.uni-bielefeld.de}
\affiliation{Bielefeld University, Thin Films and Physics of Nanostructures, Germany}
\author{Xinli Kou}
%\affiliation{Bielefeld University, Thin Films and Physics of Nanostructures, Germany}
\affiliation{Lanzhou University, School of Physical Science and Technology, Lanzhou, China}
\author{Karsten Rott}
\affiliation{Bielefeld University, Thin Films and Physics of Nanostructures, Germany}
\author{Andy Thomas}
\affiliation{Bielefeld University, Thin Films and Physics of Nanostructures, Germany}
\author{G\"unter Reiss}
\affiliation{Bielefeld University, Thin Films and Physics of Nanostructures, Germany}

\begin{abstract}
Ultra-thin magnetic tunnel junctions with low resistive MgO tunnel barriers are prepared to examine their stability under large current stress. The devices show magnetoresistance ratios of up to 110\,\% and an area resistance product of down to 4.4\,$\rm \Omega\mu m^2$. If a large current is applied, a reversible resistance change is observed, which can be attributed to two different processes during stressing and one relaxation process afterwards. Here, we analyze the time dependence of the resistance and use a simple model to explain the observed behavior. The explanation is furter supported by numerical fits to the data in order to quantify the timescales of the involved phenomena.
\end{abstract}

% 68.35.Dv Diffusion surfaces
% 72.25.-b Spin polarized transport
% 73.40.Gk Tunneling
\pacs{68.35.Dv, 72.25.-b, 73.40.Gk}

\maketitle

\section{Introduction}
The switching of a ferromagnet's magnetization by injection of a spin polarized current (Spin Transfer Torque STT switching) is an important new phenomenon in magnetism \cite{Slonczewski1996,Berger1996}. It enables applications such as spin torque nano-oscillators (STNO), Magnetic Random Access Memory (MRAM), programmable magnetic logic and sensors \cite{jmmm2008V320S1217} using the tunneling magnetoresistance (TMR) effect \cite{PRL1995V74S3273,MMM1995V139S231,prb2001V63S054416}. The conventional switching scheme of field pulses generated by current lines requires large currents and is not scalable, which led to alternatives such as heat assisted \cite{JOPcm2007V19S165218} or the already mentioned STT switching \cite{Katine2000}. In state-of-the-art MgO based magnetic tunnel junctions (MTJs), STT requires current densities of about $\rm{1 \times 10^6 A/cm^2}$ \cite{IBMjrd2006V50S81}. This leads to a bias voltage of about $\rm 100\,mV$, if the area-resistance products $RA$ are in the order of $\rm 10\,\Omega \mu m^2$.

Individual STNOs are strongly limited in their output power. This deficiency has to be overcome in order to exploit the benefits of current-tunable narrow-linewidth microwave generation for future application. Although an array of coupled oscillators may offer a path to higher power output \cite{nature2005V437S389}, the individual STNOs are envisioned to be MTJ-based and operating close to the degradation current limit \cite{jmmm2008V320S1217,nazarov2008}. Please note that STNOs are operated at direct currents in contrast to pulsed currents in the case of MRAMs.

A constant trend to reach lower area resistance products forced the fabrication of MTJs with barriers as thin as 10\,$\rm\AA$. Further reduction of the tunnel barrier thickness is becoming more and more challenging. Recently, reversible resistance changes and an atypical temperature dependence of the resistance were observed in thin $\rm AlO_x$ MTJs when reaching current densities of $10^4$ -- $10^6\,\rm A/cm^2$ \cite{prb2005V72S094432}. These resistance changes do not depend on the relative orientation of the magnetic layers \cite{prb2005V72S094432} and are attributed to electromigration in nano-constrictions of the insulating barrier \cite{mmm2005V290S1067,ieee2006V5S142}. Two opposite relaxation processes were found suggesting two independent mechanisms acting simultaneously inside the MTJs \cite{mmm2005V290S1067}. 

Here, we study the effect of current stressing on ultra-thin MgO barriers. In these devices, resistance changes take place at a current density of about $10^{6}\,\rm A/ cm^2$. This shows that further reduction of the STT switching currents is crucial in order to achieve reliable operation of STT-MTJ based devices. We distinguish two  effects during the current stress and one relaxation process. Furthermore, numerical fits to the experimental data are presented.

\section{Results}
The TMR systems with low resistive MgO barriers is sputter deposited in a Singulus TIMARIS II tool. The film sequence is Ta 3/Cu-N 90/Ta 5/Pt-Mn 20/Co-Fe 2/Ru 0.75/ Co-Fe-B 2/MgO 1.3/Co-Fe-B 3/Ta 10/Cu 30/Ru 7 (all values in nm). The composition of the compounds is: $\rm Pt_{37}Mn_{63}, Co_{70}Fe_{30}$, and $\rm Co_{66}Fe_{22}B_{12}$. Elliptical TMR elements with sizes in the range of $\rm 0.018\,\mu m^2$ to $\rm 0.095\,\mu m^2$ are prepared by e-beam lithography in combination with ion beam etching. By applying different Ar pressures during MgO deposition two sample sets with an area-resistance product of $\rm4.4\,\Omega\mu m^2$ (set 1) and $\rm9.7\,\Omega\mu m^2$ (set 2) are produced.

All measurements are performed using a constant voltage source at room temperature. By applying a voltage in the range of $\rm 200\,mV$ to $\rm 800\,mV$ the samples are stressed by a current density of 1 -- 10$\rm\times 10^6\,A/cm^2$. A magnetic field of $\rm\pm400\,Oe$ provides a stable parallel/ anti-parallel orientation of the magnetic layers during the measurements. The resistance is monitored at a read-out voltage of $20\rm\,mV$ with a rate of $\rm 1\,Hz$ in order to overcome non-linear $I(V) $-contributions and thermal drifts related with heating processes (heating rates due to current load have typical time constants well below 1\,ns \cite{jpd2007V40S5819}). Each measurement is performed on an individual element which was not stressed before. This is necessary since the properties of the barrier may change irreversibly by the stressing.

%1
\begin{figure}
 	\includegraphics{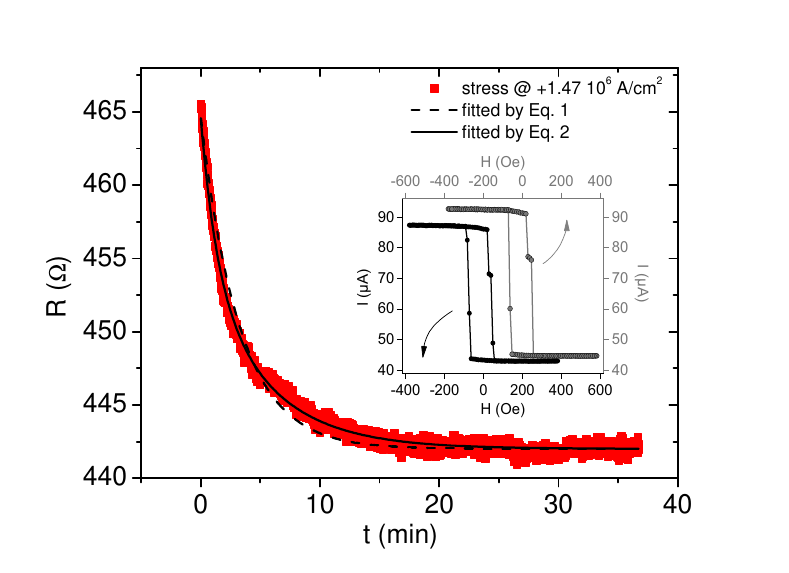} 
	\caption{Resistance as a function of time for the positive current polarity. A fast resistance decrease can be observed during the first 5\,min. The data is fitted using Eq. \ref{ansatz} and Eq. \ref{ansatz_pm} as shown by the dashed and solid line, respectively. The inset shows the minor loop of the investigated MTJ before (black) and after (gray) stressing.} 
	\label{R-t_+}
\end{figure}
%1
Figure \ref{R-t_+} shows the MTJ resistance upon stressing with positive voltage polarity (electrons flow from the bottom to the top electrode). The area-resistance product is $RA=9.7\rm\,\Omega \mu m^2$. The inset shows the magnetic minor loops before (black) and after (gray) stressing exhibiting TMR ratios of the junction of 104\,\% and 108\,\% at 20\,mV, respectively. This increase in TMR gives evidence, that the barrier itself does not degrade during current stressing. As the device is stressed by $j= +1.5\times10^6\rm\,A/cm^2$ one can see a decay of the resistance with time, although the films had been annealed for $\rm 1.5\,h$ at $\rm 360^\circ C$ prior to patterning. The initial resistance of $R_{\rm ini}=465\rm\,\Omega$ is reduced reaching a saturation resistance of $R_{\rm sat}=442\rm\,\Omega$ within $\rm 30\,min$. Numerically, the measured time dependence of the resistance can be fitted with moderate quality by an exponential function:
\begin{equation}\label{ansatz}
    R(t)=R_{\rm sat}+\Delta R\cdot\exp\left(-t/\tau\right)
\end{equation}
with $\Delta R=R_{ini}-R_{sat}>0$. This would represent a process characterized by a decay time of $\tau = \rm 3.15\,min$ and leads to the dashed line in Fig. \ref{R-t_+} as will be further discussed below.

%2
\begin{figure}
    \includegraphics{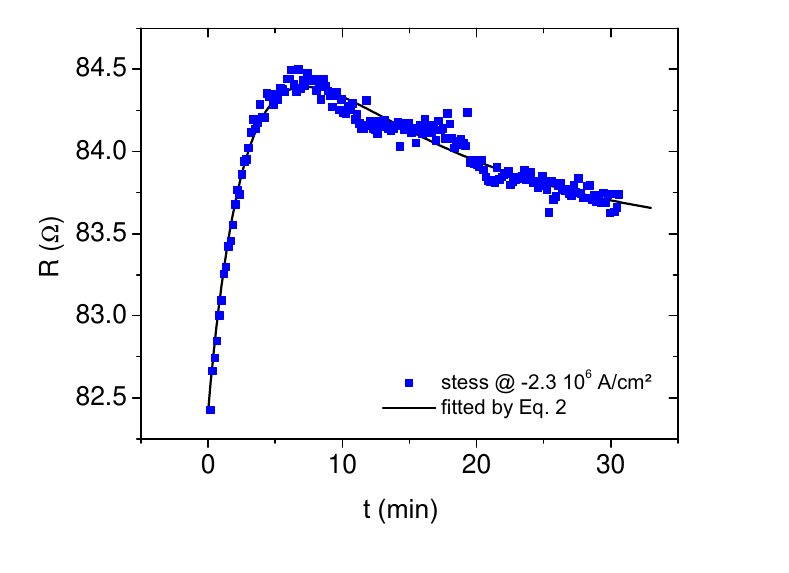} 
    \caption{Resistance as a function of time for the negative polarity. A fast increase of the resistance is followed by a slow decrease. The data is fitted by Eq. 
\ref{ansatz_pm} as shown by the solid line.}
    \label{R-t_-}
\end{figure}
%2
Stressing the MTJs with currents of the negative polarity leads to a substantially different response shown in Fig. \ref{R-t_-}. First, the resistance rapidly increases from $R_{\rm ini}=82.5\rm\,\Omega$ to $R_{\rm max}=84.5\rm\,\Omega$, followed by a slow decay. No saturation of the resistance can be seen within the measured time range.

Thus at least two processes are involved: A fast, polarity dependent resistance change followed by a slow resistance decrease for both polarities:
\begin{equation}\label{ansatz_pm}
	R(t)=R_{\rm sat}+\Delta R_1\exp(-t/\tau_1)+\Delta R_2\exp(-t/\tau_2)
\end{equation}
with $\Delta R_1>0$ ($<0$) for positive (negative) polarity and $\Delta R_2>0$ for both polarities. These two exponential decays of different sign give a very good fit to the data of Fig. \ref{R-t_-} (solid line) with $\tau_1\rm = 2.8\,min \pm 0.2\,min$ and $\tau_2\rm = 15.1\,min \pm 2.7\,min $. Looking back to Fig. \ref{R-t_+} the fast process might be hidden in the decrease because the sign depends on the polarity of the applied voltage: Two exponential decreases would lie on top of each other. In fact with Eq. \ref{ansatz_pm} a better fit to the data of Fig. \ref{R-t_+} can be obtained (solid line). In this case, however, the fit parameters $\tau_1$ and $\tau_2$ are strongly correlated and can not be used for a quantitative analysis.

%3
\begin{figure}
    \includegraphics{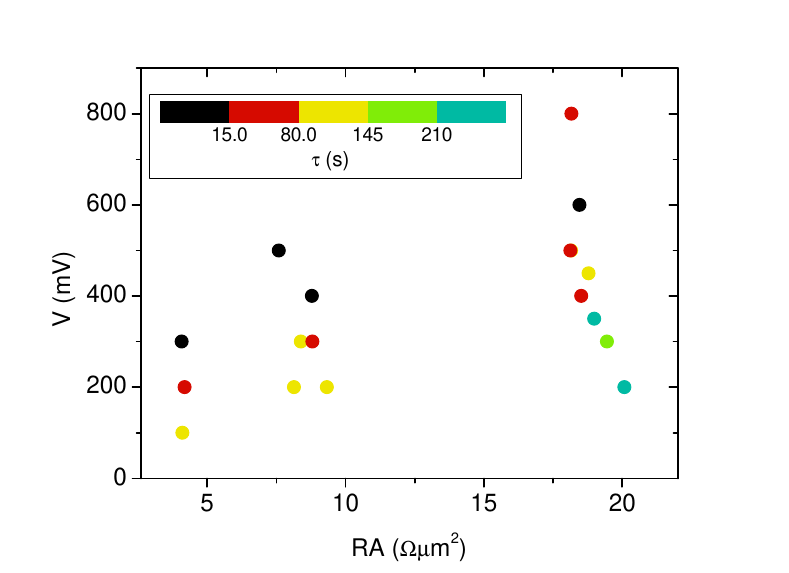} 
    \caption{Decay time $\tau$ presented by a color scale as a function of voltage and area-resistance. A decrease with increasing voltage is observed. Increasing the area-resistance at a given voltage leads to a larger $\tau$.}
    \label{t0-jV}
\end{figure}
%3
The dependence of the time constant $\tau$ (Fig. \ref{R-t_+}, dashed line) on the $RA$ product of the junctions is shown in Fig. \ref{t0-jV} for different positive stressing voltages. For increasing voltage a decrease in $\tau$ is observed. At the same time, for a given voltage, $\tau$ increases with increasing $RA$ product.

%4
\begin{figure}
    \includegraphics{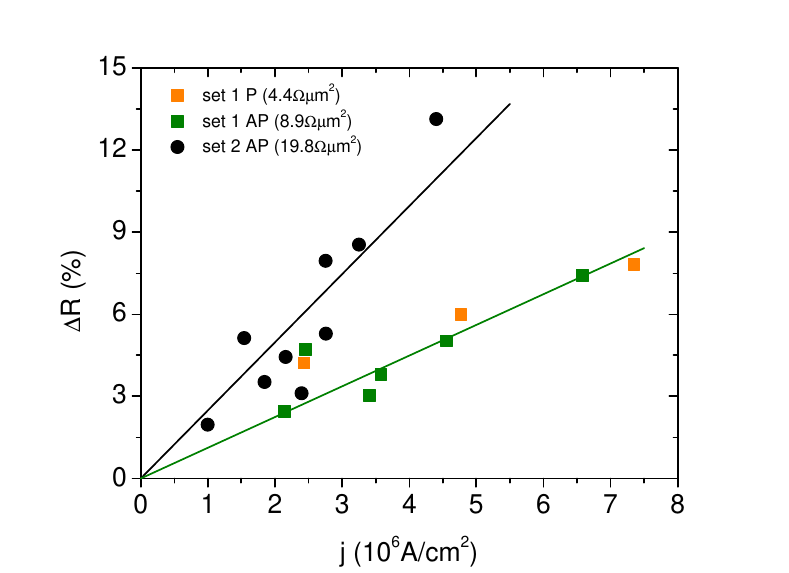} 
    \caption{Resistance change as a function of current density. The black line represents a linear fit to the results obtained from sample set 2. The slope of the green line is calculated from the difference in $RA$ between set 1 and 2. Parallel and antiparallel alignment of the magnetic electrodes is indicated by P and AP, respectively.}
    \label{dR-jV}	
\end{figure}
%4
Figure \ref{dR-jV} shows the current induced resistance variation $\Delta R$ as a function of the current density. A decrease of the resistance can be seen reaching 13\,\% at a current density of $j=4.5\times10^6\,\rm A/cm^2$ for sample set 2. For sample set 1 $\Delta R$ is reduced by a factor of approximately 0.45 due to the lower $RA$. This can be seen from Fig. \ref{dR-jV} where the black line shows a linear fit to the data of sample 2 and the green line is generated by changing the slope of the black line by the $RA$ ratio which is $8.9\,\rm\Omega \mu m^2$ to $19.8\,\rm\Omega \mu m^2$ for the antiparallel state. In contrast to this, the magnetic orientation of the electrodes (compare green and orange squares) produces no significant differences. This leads to the conclusion that the current density and not the electric field is the main parameter that determines $\Delta R$.

%5
\begin{figure}
    \includegraphics{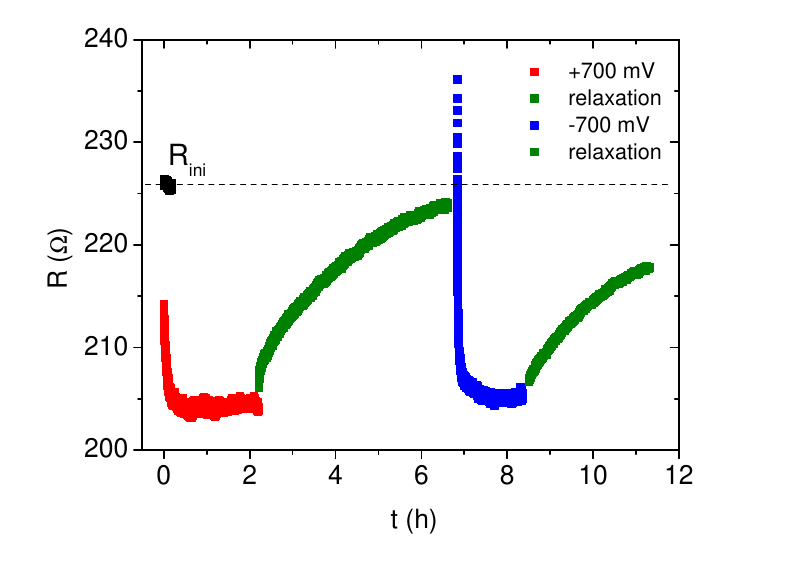}
    \caption{Resistance behavior under stressing with positive polarity (red), during the relaxation (green), stressing with negative polarity (blue), and relaxation (green). The resistance $R_{ini}$ as obtained prior to the stressing is indicated by the dashed line.}
    \label{R-t_beide} 
\end{figure}
%5
If we investigate the relaxation of the resistance, i.\ e.\ the resistance change after the current stress is released, one can reveal a third process involved. The red curve in Fig.\ \ref{R-t_beide} is equivalent to the plot shown in Fig.\ \ref{R-t_+}, the blue curve is equivalent to the plot in Fig.\ \ref{R-t_-}. The green curve in Fig.\ \ref{R-t_beide} illustrates the relaxation process: If no large voltage/ current is applied (20\,mV), the resistance slowly reaches the initial level $R_0$. This can also be fitted with an exponential function with typical relaxation times of 100\,min.

\section{Discussion}
For the interpretation of the data, one has to consider that the resistance of a (magnetic) tunnel junction is governed by the barrier thickness $d$ and the barrier height $ \Phi$. This relationship can be described by Brinkman's equation \cite{JAP1970V41S1915}. For barrier height (0.5 -- 3.5\,eV) and thickness (1 -- 1.5\,nm) that correspond to our junctions, the Brinkman equation can be reduced to 
\begin{equation}\label{brinkman}
    R(d,\Phi)=c_1\cdot\exp(c_2d\Phi^{1/2}),
\end{equation}
where $c_1$ and $c_2$ are constants. Thus, the resistance changes as given by Eq. \ref{ansatz} could be explained by a variation of $d$ as well as $\Phi$.

If we assume that the effective barrier thickness changes by a certain fraction with a nominal thickness of $\Delta d$, Brinkmans's equation leads to $\rm \Delta d\approx 0.1\,\AA$, if the resistance change is about 5\,\%. For a change of $\Phi$, a resistance change of 5\,\% leads to $\Delta \Phi \approx 10\,\rm meV$. Thus very small changes in the barrier characteristics can lead to the observed resistance changes. Provided small changes of $d$ or $\Phi$ results in a linear approximation of Eq. \ref{brinkman} and the experimental findings (Eq. \ref{ansatz}) lead to an exponential function for either $d(t)\propto\left(1+\exp(-t/\tau)\right)$ or $\Phi(t)\propto(1+\exp(-t/\tau))$ as well.

Since the layer stack is symmetric (CoFeB/ MgO/ CoFeB), the same resistance change for positive and negative bias is expected. The symmetry break is induced by the sequence of preparation. The lower ferromagnet is more likely to be contaminated with oxygen \cite{apl2007V90S132503}, because the MgO is deposited on top of the ferromagnet by rf-sputtering from an MgO target. This oxygen on top of the ferromagnet can act as an additional barrier \cite{prb2003V68S092402} and also lead to oxygen deficiencies in the lower part of the tunnel barrier. Also the effective barrier height $\Phi$ is modified by the presence of oxygen vacancies \cite{prb1994V50S2582,prb2006V73S205412}.

Therefore, a possible explanation for the observed behavior is a small displacement of these oxygen vacancies/ defect atoms driven (fast) in the direction of the electron flow increasing or decreasing the effective thickness of this layer (electromigration) \cite{ieee2006V5S142} and (slow) by thermal activation. These processes are reversible since the atoms can slowly fall back into their initial positions of lower energy once the current is switched off.

\section{Summary}
In summary, we investigate magnetic tunnel junctions with MgO barriers with magnetoresistance ratios of up to 110\,\% and area-resistances of around 10\,$\rm\Omega \mu m^2$. The low resistance of these devices allow us to drive large currents trough the structures, which is necessary for use in spin transfer torque memory or logic cells. We observe three reversible resistance changes of the junctions: One fast, polarity dependent term that could be associated with electromigration. One slower, polarity independent term active during current stress, that is suggested to be of thermal origin. Finally, a very slow relaxation takes place when the current stress is released, which also can be explained by thermally activated diffusion.

\section*{Acknowledgments}
The authors gratefully acknowledge Singulus Technologies for providing the samples. P.~K.\ is supported by the Deutsche Forschungsgemeinschaft (DFG) through Schwerpunktprojekt 1133. X.~K.\ is supported by the DFG within Grant No. RE\,1052/16-1.

%\bibliography{datenbank}

\newpage

\end{document}